\def\ga5{\gamma_5}

\def\r{\mbox{\boldmath r}}
\def\L{{\cal{L}}}
\def\tr{\mbox{tr}}

\def\tr {\, \mbox{tr} \,}
\def\Trp{\, \mbox{Tr}^{\prime} \,}
\def\trp{\, \mbox{tr}^{\prime} \,}

\def\Mt {{\cal{M}}^\theta}
\def\S  {{\cal{S}}}
\def\L  {{\cal{L}}}

\def\Trp{\, \mbox{Tr}^{\prime} \,}
\def\r{\vec{r}}
\def\al{\alpha}
\def\square{\hbox{\vrule\vbox{\hrule\phantom{o}\hrule}\vrule}}
\documentstyle[12pt]{article}
\title{ \Large
                   Chiral Meson Lagrangians from the QCD Based \\
                   NJL Model Modified by Nonlocal Effects}
\author{
A.A.Bel'kov${}^1$,
A.V.Lanyov${}^1$,
A.Schaale${}^{2*}$,
S.Scherer${}^2$
\\
\\
\small
${}^1$
        Particle Physics Laboratory, Joint Institute for Nuclear
Research,
\hfill\\
\small
        141980 Dubna, Moscow Region, Russia
\hfill\\
\small
${}^2$
  TRIUMF, 4004 Wesbrook Mall, Vancouver, B.C., Canada V6T 2A3
\hfill\\
\small
${}^*$ (supported by Deutscher Akademischer Austauschdienst, DAAD)
\hfill
\\
\\  TRI-PP-94-19
}

\begin{document}

\thispagestyle{empty}
\begin{titlepage}
\thispagestyle{empty}
\maketitle
\begin{abstract}

Starting from a QCD inspired bilocal quark interaction we obtain a
local effective meson lagrangian. In contrast to previous local
(NJL-like) approaches, we include nonlocal corrections related to
the finite meson size which we characterize by a small parameter.
After bosonization using the heat-kernel method we predict the
structure coefficients of the Gasser-Leutwyler $p^4$-lagrangian
up to first order in this parameter. The modifications for the
$L_i$ coefficients are typically of the order 15-20\%, except for
$L_5$, where we find a stronger nonlocal influence.

\end{abstract}
\end{titlepage}

%--------------------------------------------------------------------
\section{Introduction}
%--------------------------------------------------------------------

    Quantum chromodynamics (QCD) is generally accepted as the theory
of strong interactions in terms of quarks and gluons, and is
experimentally well
established in the perturbative regime at high energies $(E\gg1\,\mbox{GeV})$.
    Since quarks and gluons have not been observed as free particles,
it is assumed that they are confined into color-neutral hadrons.
Many phenomenological approaches and effective models exist in the
literature to describe hadrons and their interactions at low and medium
energies $(E<1\,\mbox{GeV})$.
    However, up to now it is still impossible to derive a meson
theory from QCD in a mathematical exact way, including the confinement
of quarks and gluons.
    This problem is connected, in particular, with the unknown behavior
of the QCD Green's functions at low energies.
    Nevertheless, there is some success related with effective
bilocal approaches based on the application of functional methods to
approximate forms of QCD
(see \cite{kleinert}--\cite{ball2} and references therein).
In the local limit, these approaches reduce to the Nambu--Jona-Lasinio
(NJL) model \cite{njl}. The bosonization of the NJL model leads to
effective chiral meson lagrangians \cite{ebert-reinhardt}--\cite{bijnens}.

    In this work we derive an effective chiral $p^4$-lagrangian of the type
established by Gasser and Leutwyler \cite{gasser} using the effective
bilocal approach. We introduce a small parameter $\al$ characterizing
the size of the nonlocal
corrections. We predict the structure constants of the effective
$p^4$-lagrangian to first order in $\al$.
Of course, in the limit $\al \to 0$, our method reproduces the
results of the NJL model.
In Section 2 we briefly review the standard method of transforming
the QCD lagrangian into an effective 4-quark lagrangian.
In Section 3 we estimate the size of the $\overline{q}q$ system within a
nonrelativistic Schr\"odinger approach for constituent quarks.
We use the result for a quantitative estimate of the nonlocal
effects and then describe a fixed-distance approximation.
In section 4 we present the bosonization in the more
general dynamical bilocal approach. For that purpose we consider a
separable ansatz for the bilocal collective fields.
Solving the heat-kernel equation, in Section 5 we calculate the
modified structure constants $L_i$ of the chiral $p^4$-lagrangian.

%--------------------------------------------------------------------------
\section{From QCD to the Bilocal Effective Action}
%--------------------------------------------------------------------------

 The starting point of our consideration is the generating functional of
QCD in Minkowski space,
\begin{equation}
\label{genfunc}
Z[\xi, \overline{\xi},J^a_{\mu}]
\;=\;\int D\overline{q}Dq DA\; \mbox{exp} \bigg(i {\S} [\overline{q}, q, A]
     + i \int d^4x \big(\overline{q}\xi+\overline{\xi}
q+J^a_{\mu}A^{a\,\mu}\big) \bigg)\,,
\end{equation}
where
\begin{equation}
{{\S}} [{\overline{q}} ,q,A]\;=\;\int d^4 x
\bigg[ \overline{q}(i \hat{D} - m_0)q
\;-\; \frac{1}{4} \sum_{a=1}^8 G_{\mu\nu}^a G^{a\,\mu\nu} \bigg]\,,
\label{sqcd}
\end{equation}
and $\xi$, $\overline{\xi}$, $J^a_{\mu}$ are the external sources associated
with
the fields $\overline{q}$, $q$, $A^a_{\mu}$; $q$ is the quark field; $m_0$ is
the current
quark mass matrix; $A^a_\mu$ represents a gluon with color index $a$, and
$\lambda^a_C$ are $SU(3)_C$ matrices.
    The covariant derivative is defined as
\begin{equation}
\label{covder}
D_{\mu}=\partial_{\mu}-ig \sum_{a=1}^{8} \frac{\lambda^a_C}{2} A^a_{\mu}\,,
\end{equation}
and the gluon field-strength tensor is of the form
\begin{equation}
G_{\mu\nu}^a=\partial_\mu A_\nu^a - \partial_\nu A_\mu^a
                - gf_{abc} A_\mu^b A_\nu^c\,,
\end{equation}
where $g$ is the QCD coupling constant, and $f_{abc}$ are the $SU(3)$
structure constants.
We use $\hat{D}$ for $\gamma_{\mu} D^{\mu}$.
The Faddeev-Popov ghost fields and gauge fixing terms are included in
the gluon measure.

    Using standard techniques of path integration
\cite{kleinert}--\cite{cahill1}, after integration over the gluon
fields, one finds
\begin{eqnarray}
Z \;=\; \int D\overline{q}D q \;\mbox{exp} \bigg[i \int d^4 x
\;\overline{q}(x)(i
\hat{\partial} - m_0) q(x)\bigg] \; \mbox{exp} \big(iW[j] \big) \; ,
\label{z2}
\end{eqnarray}
where we drop a normalization factor and do not consider external quark
sources.
In Eq.\ (\ref{z2}) $W[j]$ is given by
\begin{equation}
W[j] \;=\; \sum_{n=2}^{\infty} \frac{1}{n!}
\int d^4x_1...d^4x_n D^{a_1...a_n}_{\mu_1...\mu_n}(x_1,...,x_n)
\prod_{i=1}^{n} j^{a_i}_{\mu_i}(x_i)\;,
\label{W}
\end{equation}
where
$$
j^a_{\mu}(x) \;=\; \bar{q}(x) \gamma_{\mu} \frac{\lambda_C^a}{2} q(x)
$$
is the flavor-singlet local quark current.
    The function $D^{a_1...a_n}_{\mu_1...\mu_n}$ is the $n$-point connected
gluon Green's function containing all the information about the gluon dynamics.
    As long as the behavior of the gluon propagator and the running
coupling constant at long distances are unknown,
an analytical integration of Eq.\  (\ref{W}) is
impossible.

    If one retains only the first term in the expansion of $W[j]$ of
Eq.\  (\ref{W}),
neglecting  triple and higher-order gluon vertices, the generating functional
can be written in the effective form of truncated QCD,
\begin{equation}
Z \;=\; \int D\overline{q}Dq \;
 \mbox{exp} \Big\{ i\Big[ \int d^4x\; \bar{q}(x) \big(
i\hat{\partial} - m_{0} \big)q(x)\Big] + i{\S}_{int} \Big\}\,,
\label{zz}
\end{equation}
where
\begin{equation}
{\S}_{int} \;=\; - i{{g^2}\over{2}} \int \hspace{-0.2cm}\int d^4x\,d^4y\;
 j^{a\,\mu}(x) \, D_{\mu\nu}^{ab}(x-y) j^{b\,\nu}(y)
\label{sint}
\end{equation}
is the part of the effective action corresponding to quark interaction
via single-gluon exchange. In the Feynman gauge the nonperturbative gluon
propagator
is defined as
\begin{equation}
  D_{\mu\nu}^{ab}(x) = g^{-2} \delta^{ab} g_{\mu\nu} D(x)\,.
\label{gluon}
\end{equation}

As  the behavior of the Green's function $D(x)$ is unknown for large
distances, a specific ansatz has to be used.
For example, after a Fierz transformation, the local ansatz
$D(x)\sim\delta^{(4)}(x)$ will lead to the NJL-type lagrangian of the effective
four-quark interaction \cite{njl},
\begin{equation}
\L_{int}= 2G_1\bigg[ \bigg(\overline{q}\frac{\lambda^{a}}{2} q \bigg)^2
                   +\bigg(\overline{q}i\gamma_5 \frac{\lambda^{a}}{2}q\bigg)^2
\bigg]
        -2G_2\bigg[ \bigg(\overline{q}\gamma_\mu \frac{\lambda^{a}}{2}q
\bigg)^2
                   +\bigg(\overline{q}\gamma_5\gamma_\mu
\frac{\lambda^{a}}{2}q\bigg)^2
             \bigg]\, ,
\label{NJL}
\end{equation}
with some universal coupling constants $G_1$ and $G_2$.
    In the following  we discuss an approach which goes beyond the usual
local ansatz.

%--------------------------------------------------------------------------
\section{Bilocal Fixed-Distance Approximation}
%--------------------------------------------------------------------------

   Here we will outline the ideas which motivate the bilocal
{\em fixed-distance} approach.
   One of the essential principles of a bilocal meson theory is the choice of
the relativistic covariant form for the instantaneous four-quark
interaction.
   The existence of this interaction type is the direct consequence of
the reduced phase-space method of quantization of chromodynamics
\cite{pervushin}. Assuming  the dominant role of equal-time
interactions in the formation of bound states, the Schwinger-Dyson
equations of the bilocal meson theory reduce to the nonrelativistic
Schr\"odinger approach corresponding to the description of
$\overline{q}q$-pairs
interacting via some effective gluonic potential.

   If a meson is considered as a bound $\overline{q}q$-system analogous to the
hydrogen atom in quantum mechanics, the nonrelativistic Schr\"odinger
equation reads \cite{quigg-rosner} ($\hbar=c=1$)
\begin{equation}
-\frac{1}{2m} \nabla^2 \Psi(\r) + [V(\r)-E]\Psi(\r)=0\,.
\label{sch1}
\end{equation}
In Eq.\ (\ref{sch1}) $\Psi$ is the wave function of the internal motion, $m$ is
the
reduced constituent quark mass
of a two-body system, $m=(m_1 m_2)/(m_1+m_2)$, ${\r}$ is the relative
coordinate, $V(\r)$ is the interaction potential and $E$ is the eigenvalue of
the hamiltonian. For a spherical potential
    the wave function $\Psi$ is usually written as a product of the
radial function $R(r)$ with $r=|\r|$ and the spherical harmonics
$Y_{lm}(\Theta,\phi)$: $\Psi(\r)=R(r)Y_{lm}(\Theta,\phi)$.
    The radial Schr\"odinger equation is of the form
\begin{equation}
- u''(r) = 2m \Big[E-V(r)-\frac{l(l+1)}{2mr^2}\Big] u(r)\;,\; u(r)=rR(r)\;,
\label{sch2}
\end{equation}
with the boundary condition $u(0)=0$.

    The results of a QCD lattice analysis \cite{laermann} show that at
large distances the
effective quark-antiquark interaction can be approximated by a linear
potential
\footnote{Screening effects by virtual $\overline{q}q$ pairs at very large
distances
          are omitted here.}.
If one approximates the quark--antiquark potential by a linear
potential
(Coulomb and other corrections are neglected here),
\begin{equation}
V(r)=\sigma\cdot r,
\label{pot}
\end{equation}
with $\sigma \approx 0.27\; GeV^2$ for $m_u=m_d=\mu=0.336\;GeV$
\cite{lucha}, the characteristic distance between the quark and
antiquark, $<r>$, can easily be determined using the virial theorem,
\begin{equation}
 <r>\equiv h=\frac{2 E_1}{3\sigma}\approx 0.68\, fm ,
\label{bohr}
\end{equation}
where $E_1=2.238 (\sigma^2/2\mu)^{1/3}$ is the ground state energy.
In the following we only consider the ground state ($l=0$), i.e., neglect
excited
mesonic states such as $\pi^*$, $K^*$ etc..
After scaling Eq.\  (\ref{sch2}) by introducing
$\rho=r/h_0$ and $w(\rho)=u(r)$ one finds $ w''(\rho) +(\varepsilon -
2\rho)w(\rho)=0$, with $\varepsilon=2m(2m\sigma)^{-\frac23} E$.
    For large $\rho$ the solution behaves like the Airy function,
$w(\rho)\sim Ai(\rho)$, decreasing exponentially to zero.
    The radial wave function $R(r)$ defined in Eq.\ (\ref{sch2}) decreases
even stronger.
Thus one expects a small root-mean-square deviation
$\Delta r$ for the distance between the constituent quarks.

In order to obtain a qualitative estimate of the nonlocal
corrections from the  bilocal effective action, we make the following
ansatz. We consider the case where the constituent quarks in the meson
are localized at the scale $h$.
  This we do by including a delta function $\delta((x-y)^2-h^2)$
into the integrand $S_{int}$, Eq.\ (\ref{sint}),
\begin{eqnarray}
{\S}_{int} &=& -i \frac{\kappa^2}{2} \int \hspace{-0.2cm}\int d^4x \, d^4y \;
    j^a_{\mu}(x) \, j^{a\,\mu}(y) \, D(x-y) \;
\delta\left((x-y)^2-h^2\right)\,,
\label{sh1}
\end{eqnarray}
where the correct dimension is obtained by introducing a constant $\kappa$
($\;([\kappa]=m^{-1}$).
After shifting the argument $y$ by a Lorentz-invariant operator,
$$
q(y)=\mbox{exp}\Big((y-x)_\mu \partial^\mu \Big)\;q(x)\;,
$$
the effective action, Eq.\ (\ref{sh1}), becomes
\begin{equation}
{\S}_{int} = -i \frac{\kappa^2 D(h)}{2} \int d^4x \;
              j^a_{\mu}(x) K(h,x)\;
              j^{a\,\mu}(x) \,,
\label{sh2}
\end{equation}
where
\begin{equation}
K(h,x)= \int d^4y \;\mbox{exp}\big((y-x)_\mu \partial^\mu\big)\;
\delta\big((x-y)^2-h^2\big)\;.
\label{kx1}
\end{equation}
Performing the integration in polar coordinates, Eq.\  (\ref{kx1})
can be expanded in the following way,
\begin{equation}
K(h,x) = \pi^2 h^2 \sum_{n=0}^{\infty} \frac{1}{(2n)!}
         \frac{\Gamma(n+\frac{1}{2})}{\Gamma(\frac{1}{2})\Gamma(n+2)} h^{2n}
         \Box^n
       = \pi^2 h^2 \bigg[1+\frac{1}{8}\frac{\Box}{\Lambda^2}
         + O\bigg(\frac{\Box^2}{\Lambda^4}\bigg)\bigg]\,,
\nonumber
\end{equation}
where $\Lambda=h^{-1}$, and thus Eq.\ (\ref{sh2}) can be
written in the form
\begin{equation}
{\S}_{int} = -i \frac{9G}{16} \int d^4x \; \bigg[
                j^a_{\mu}(x) \;
                j^{a\,\mu}(x)
                +\frac{1}{8\Lambda^2}
                j^a_{\mu}(x) \left( \Box j^{a\,\mu}(x) \right)
                                              \bigg]
               +O\bigg(\frac{\Box^2}{\Lambda^4} \bigg)\;,
\label{sh3}
\end{equation}
with $G=\frac{8}{9}\pi^2 \kappa^2 h^2 D(h)$.

    After a Fierz transformation the action, Eq.\ (\ref{sh3}), reads
\begin{eqnarray}
{\S}_{int} &=& i \frac{9G}{16} \int d^4x \; \bigg(
                 \overline{q}(x) \frac{\Mt}{2}
q(x)\;\overline{q}(x)\frac{\Mt}{2} q(x)
\nonumber \\
             && +\frac{1}{8\Lambda^2}
                 \overline{q}(x)\frac{\Mt}{2}\;\Box
\big[q(x)\overline{q}(x)\big]\frac{\Mt}{2}q(x)
                                                \bigg)\,,
\label{Sint}
\end{eqnarray}
where ${\Mt}$ are tensor products of Dirac, flavor and color
matrices of the type
$$
   \bigg\{ {\bf 1}\,, i\gamma_5\,,
i\sqrt{\frac{1}{2}}\gamma^{\mu}\,,
          i\sqrt{\frac{1}{2}}\gamma_5\gamma^{\mu} \bigg\}^D
   \bigg\{ \frac{1}{2}\lambda^a_F \bigg\}^F
   \bigg\{ \frac{4}{3} {\bf 1} \bigg\}^C \,.
$$
Here we consider the $SU(3)_F$ flavor group with flavor matrices $\lambda^a_F$,
and we restrict ourselves to the color-singlet $\bar{q}q$ contributions.
    The first term in Eq.\ (\ref{Sint}) leads to the effective four-quark
interaction of the NJL model, Eq.\ (\ref{NJL}), with $G_1=2G_2=G/4$ while the
second term, proportional to $1/\Lambda^2$,
takes into account the finite-size effects of collective mesons.

%---------------------------------------------------------------------------
\section{Dynamical Bilocal Approach}
%---------------------------------------------------------------------------

    Before considering the physical results of the bilocal fixed-distance
approximation, let us study a more general
approach using a {\em dynamical} bilocal model.
    Starting with the effective action, Eq.\ (\ref{sint}), and performing
a Fierz transformation one finds
\begin{equation}
S_{int} \;=\; {{i}\over{2}} \int \hspace{-0.2cm}\int
d^4x\,d^4y\;
D(x-y)\,\bar{q}(x) \frac{\Mt}{2} q(y)\,
\bar{q}(y) \frac{\Mt}{2}q(x)\,.
\label{fsint}
\end{equation}
Introducing scalar $(S)$, pseudoscalar $(P)$, vector
$(V)$ and axial-vector $(A)$ bilocal collective meson fields
\cite{kleinert}--\cite{cahill1}  leads to an effective
action which is bilinear in the quark fields,
\begin{eqnarray}
S_{int} &=& \int \hspace{-0.2cm}\int d^4x\,d^4y\; \bigg\{
   -\frac{9}{8D(x-y)} \tr \bigg[ \big(\widetilde{S}(x,y) \big)^2
                               +\big(\widetilde{P}(x,y) \big)^2
\nonumber \\
&& +2\Big(\big(\widetilde{V}_{\mu}(x,y) \big)^2
       +\big(\widetilde{A}_{\mu}(x,y) \big)^2
     \Big) \bigg]
   + \overline{q}(x)\widetilde{\eta}(x,y)q(y) \bigg\}\; ,
\label{Sclfld}
\end{eqnarray}
with
\begin{equation}
\widetilde{\eta}(x,y) =
-\widetilde{S}(x,y) - i \gamma^5 \widetilde{P}(x,y)
+ i \gamma^{\mu} \widetilde{V}_{\mu}(x,y)
+ i \gamma^{\mu} \gamma^5 \widetilde{A}_{\mu}(x,y) \; ,
\label{eta}
\end{equation}
where
\begin{equation}
    \widetilde{S}=\widetilde{S}^a\frac{\lambda^a}{2} , \quad
    \widetilde{P}=\widetilde{P}^a\frac{\lambda^a}{2} , \quad
    \widetilde{V}_{\mu}=-i\widetilde{V}_{\mu}^a \frac{\lambda^a}{2}
,\quad
    \widetilde{A}_{\mu}=-i\widetilde{A}_{\mu}^a \frac{\lambda^a}{2}
\nonumber
\end{equation}
are the matrix-valued collective fields associated with the following
quark bilinears,
\begin{eqnarray}
\widetilde{S}^{a}(x,y) &=& -\frac{8}{9}D(x-y)
      \overline{q}(y)\frac{\lambda^{a}}{2}q(x)\,,\quad
\nonumber \\
\widetilde{P}^{a}(x,y) &=& -\frac{8}{9}D(x-y)
      \overline{q}(y)i\gamma^{5}\frac{\lambda^{a}}{2}q(x)\,,
\nonumber \\
\widetilde{V}_\mu^{a}(x,y) &=& -\frac{4}{9}D(x-y)
      \overline{q}(y)\gamma_\mu \frac{\lambda^{a}}{2}q(x)\,,\quad
\nonumber \\
\widetilde{A}_\mu^{a}(x,y) &=& -\frac{4}{9}D(x-y)
      \overline{q}(y)\gamma_\mu \gamma^{5}\frac{\lambda^{a}}{2}q(x)\,.
\nonumber
\end{eqnarray}

Following  ref. \cite{ball2}, we assume a strong localization of the
bilocal fields and make the ansatz
\begin{equation}
\widetilde{\eta}(x,y) \to \widetilde{\eta}(z,t) =
\eta(z)f(t) + \eta_\mu(z)t^\mu g(t) + \cdots\;,
\label{eta2}
\end{equation}
where $z=(x+y)/2$, $t=(y-x)/2$ are the global and relative
coordinates, respectively. The function
\begin{equation}
\eta(z) \;=\; - S(z) - i\gamma^5 P(z)
              + i\gamma^{\mu} V_{\mu}(z) + i\gamma^{\mu} \gamma^5 A_{\mu}(z)
\label{eta1}
\end{equation}
combines the local collective fields of the composite operators
$\overline{q}(z)q(z)$, ${}$ $\overline{q}(z)i\gamma^{5}q(z)$,
$\overline{q}(z)\gamma_\mu q(z)$ and
$\overline{q}(z)\gamma_\mu \gamma^{5}q(z)$, corresponding to the lowest meson
excitations $0^{++}$, $0^{-+}$, $1^{--}$, $1^{++}$.
    The next order term of Eq.\  (\ref{eta2}), proportional to
$\eta_\mu$,
can be identified with the excitations $1^{--}$, $1^{+-}$, $2^{++}$,
$2^{--}$ \cite{ball2}.
The functions $f(t)$ and $g(t)$ rapidly decrease for $|t^2|\gg h^2$ and
strongly localize the bilocal fields $\widetilde{\eta}(x,y)$
to the effective size of the collective meson
$h \equiv 1/\Lambda$.

    Expanding $q(y)$ and $\overline{q}(x)$ in a Taylor series about $z$,
$$
      q(y) \;=\; q(z) +t^{\mu}\partial_{\mu}q(z) + O(t^2)\; ,\;
      \overline{q}(x) \;=\; q(z) -t^{\mu}\partial_{\mu}q(z) + O(t^2)\; ,
$$
and using Eq.\ (\ref{eta2}) we obtain
\begin{eqnarray*}
\int \hspace{-0.2cm}\int d^4x\,d^4y\; \bar{q}(x)\widetilde{\eta}(x,y)q(y) &=&
2\int d^4z\, \bar{q}(z)\eta(z)q(z) \int d^4t\,f(t)
\nonumber
\\&+&
2\int d^4z \partial^{\mu}\bar{q}(z)\eta(z)\partial_{\mu}q(z)
            \int d^4t\,t^2f(t)
\nonumber
\\&+&
 \mbox{(excitation terms)}\,.
\end{eqnarray*}
    Then, for the first generation of mesons corresponding to the
$\big(0^{++}$, $0^{-+}$, $1^{--}$, $1^{++}\big)$ multiplets, the generating
functional is
\begin{eqnarray}
{\cal Z} &=& \int {\cal D}\Phi\,{\cal D}\Phi^\dagger\,{\cal D}V\,{\cal D}A
           \,\,\mbox{exp} \bigg\{ \int d^4 z
           \bigg[- {1 \over{4G_1}} \mbox{tr} [(\Phi(z) -
m_0)^\dagger(\Phi(z)-m_0)]
\nonumber
\\&&       -\,{1 \over{4G_2}} \mbox{tr} \big(V_\mu^2(z) + A_\mu^2(z) \big)
           +\bar{q}(z) i\widehat{\bf D} q(z)
\nonumber
\\&&       -\frac{\alpha}{\Lambda^2}
            \partial^{\mu}\bar{q}(z)\,\eta(z)\,\partial_{\mu}q(z)
           \bigg] \bigg\}\; ,
\label{NJLnonloc}
\end{eqnarray}
where ${\bf \widehat{D}}$ is the Dirac operator in the presence of
local collective meson fields,
\begin{eqnarray}
 i{\bf \widehat{D}} &=& i(\widehat{\partial}+\widehat{V}+\widehat{A}\gamma^5)
                      - P_{R}(\Phi +m_{0}) - P_{L}(\Phi^\dagger+m_{0})
\nonumber
\\
        &=& [i(\widehat{\partial} +\widehat{A}^{(+)}) - (\Phi    +m_0)] P_R
           +[i(\widehat{\partial} +\widehat{A}^{(-)}) - (\Phi^\dagger+m_0)]
P_L.
\label{dirac}
\end{eqnarray}
    Here $\Phi = S + iP$, $\widehat{V} = V_{\mu} \gamma^{\mu}$,
$\widehat{A} = A_{\mu} \gamma^{\mu}$;
$P_{R/L}={1 \over 2}(1 \pm \gamma_5)$ are chiral right/left projectors;
$\widehat{A}^{(\pm)} = \widehat{V} \pm \widehat{A}$ are right and left
combinations of fields. The parameter $\al$ is defined as
\begin{equation}
\frac{\alpha}{\Lambda^2}\;=\;\frac{1}{2}\int d^4t\,t^2f(t)\,,
\label{alpha}
\end{equation}
where $f(t)$ is normalized as $2\int d^4t f(t) =1$.

The coupling constants $G_1$ and $G_2$ are
defined by
\begin{equation}
\frac{1}{G_1} \;=\; \frac{1}{2G_2} \;=\;
                    \frac{9}{8}\int d^4t\,\frac{f^2(t)}{D(2t)}\,.
\label{ft}
\end{equation}
Note that in such an approximation the ratio of $G_2$ and $G_1$ is
$1/2$, whereas phenomenology predicts
$G_2 / G_1 \sim 4$.
This problem can, in principle, be solved by introducing different
functions $f_\sigma(t)$ $(\sigma=0,1,\cdots)$
into Eqs.\ (\ref{eta2},\ref{eta1}), corresponding to different
localizations of spin-0 and spin-1 mesons.
However, in our approximation we neglect such a spin dependence.

    The first three terms of the action of Eq.\ (\ref{NJLnonloc}) are
identical with the
corresponding terms arising from the linearization of the four-quark
local interaction of the extended NJL model described by the lagrangian
$$
\L_{NJL}\;=\;\overline{q}(i \hat{\partial} - m_0)q + \L_{int}\;,
$$
with $\L_{int}$ given by Eq.\ (\ref{NJL}).
    Performing a partial integration and dropping the surface term,
the last term in Eq.\ (\ref{NJLnonloc}) can be rewritten in the form
\begin{equation}
 \int d^4 z \partial^{\mu}\bar{q}(z)\,\eta(z)\,\partial_{\mu}q(z) \;=\;
-\int d^4 z \bar{q}(z)\big[ \partial^{\mu}\eta(z)\,\partial_{\mu}
                           +\eta(z)\,\partial^2 \big] q(z) \,.
\label{partint}
\end{equation}

Of course, we do not know the explicit form of the function $f(t)$.
However, we can now use the fixed-distance approximation of Eq.\
(\ref{sh1}) to estimate the parameter $\al$.
Indeed, the second term in Eq.\  (\ref{Sint}) can be transformed into the form
\begin{equation}
{\S}_{int}^2\;=\;\frac{i}{16\Lambda^2} \int d^4x
                         \bar{q}(x)\big[ \partial^{\mu}\eta(x)\,\partial_{\mu}
                        +\eta(x)\,\partial^2 \big] q(x)
                        +\mbox{(excitation terms)}\,,
\label{sintlam}
\end{equation}
where $\eta(x)$ is the combination of the local collective fields,
Eq.\ (\ref{eta1}),
which now are defined by the following quark bilinears,
$$
S^a(x) = -G \,\overline{q}(x)\frac{\lambda^a}{2} q(x)\,, \quad
P^a(x) = -G \,\overline{q}(x) i\gamma_5\frac{\lambda^a}{2} q(x)\,,
$$
$$
V^a_{\mu}(x)=-\frac{G}{2}\,\overline{q}(x)\gamma_\mu\frac{\lambda^a}{2}q(x)\,,\quad
A^a_{\mu}(x)=-\frac{G}{2}\,\overline{q}(x)\gamma_\mu\gamma_5\frac{\lambda^a}{2}q(x)\,.
$$
Comparing Eqs.\ (\ref{partint}) and (\ref{sintlam}) we can
fix the value  $\alpha$,  $\alpha = 1/16$, corresponding to the naive bilocal
fixed-distance approximation.
    Using the values $\Lambda = 0.28\,\mbox{GeV}$ and $\mu=0.336\,\mbox{GeV}$,
the ratio is estimated to be of the order
\begin{equation}
\frac{\alpha \mu^2}{\Lambda^2} \approx 0.09\; .
\label{ratio}
\end{equation}
This value will be used as a small parameter for further numerical estimates of
nonlocal effects.

After integration over the quark fields, the full
action arising from the generating functional, Eq.\ (\ref{NJLnonloc}), is
\begin{eqnarray}
{\S}(\Phi,\Phi^\dagger,V,A) &=& \hspace{-0.2cm}\int d^4 z
\bigg [ - {1 \over{4G_1}} \mbox{tr} [(\Phi -m_0)^\dagger(\Phi-m_0)]
           - {1 \over{4G_2}} \mbox{tr} (V_\mu^2 + A_\mu^2) \bigg]
\nonumber \\
&& -i\,\Trp [\log(i {\bf {\widehat{\widetilde{D}}}})]\,.
\label{action}
\end{eqnarray}
  Here the second term is the quark determinant of the Dirac operator
$i{\bf {\widehat{\widetilde{D}}}}$ which is extended to the case
of nonlocality. It is
obtained from the usual operator, Eq.\  (\ref{dirac}), by the following
replacement,
\begin{eqnarray}
&& A^{(\pm)}_{\mu} \to
A^{(\pm)}_{\mu}\bigg(1 +\frac{\alpha}{\Lambda^2}\partial^2 \bigg)
+
\frac{\alpha}{\Lambda^2}\Big(\partial_{\nu}A^{(\pm)}_{\mu}\Big)\,
\partial^{\nu}\,,
\nonumber \\
&& \Phi \to
\Phi \bigg(1 +\frac{\alpha}{\Lambda^2}\partial^2 \bigg)
+ \frac{\alpha}{\Lambda^2}\Big(\partial_{\nu} \Phi\Big)\,\partial^{\nu}\,.
\label{shift2}
\end{eqnarray}
The "trace" $\Trp$ is to be understood as a space-time integration
and a ``normal'' trace with respect to Dirac, color and flavor matrices,
$$\Trp = \int d^4 x \trp \,,\quad
  \trp = \tr_{\gamma} \cdot \tr_C \cdot \tr_F \,.
$$

In the following we will only consider the non-anomalous part of the
effective meson action which corresponds to the modulus of the quark
determinant
\footnote{The imaginary part of the quark determinant is
related to the anomalous part of the action.}.
    The modulus of the quark determinant can be calculated using the
heat-kernel technique with proper-time regularization
\cite{heat-other}--\cite{our},
\begin{equation}
  \log |\det i {\bf \widehat{\widetilde D}}| \,  =
- \frac{1}{2}\Trp \log ({\bf \widehat{\widetilde D}}^\dagger{\bf
\widehat{\widetilde D}}) \,  =
 - {1\over 2} \int^{\infty }_{1/\widetilde{\Lambda}^{2}} d\tau{1\over \tau}
                    \Trp \exp (-{\bf \widehat{\widetilde D}}^\dagger{\bf
\widehat{\widetilde D}\tau})\; ,
\label{logarithm}
\end{equation}
with $\widetilde{\Lambda}$ as the cutoff parameter of intrinsic regularization.
    The operator ${\bf \widehat{\widetilde D}}^\dagger{\bf \widehat{\widetilde
D}}$ can be written as
\begin{eqnarray*}
{\bf \widehat{\widetilde D}}^\dagger{\bf \widehat{\widetilde D}}
&=& \beta \partial^2 + \mu^2
  + 2\Gamma_{\mu}\partial^{\mu} + \Gamma_{\mu}^2 + a(x)
\nonumber \\
  &+& \frac{\alpha}{\Lambda^2}\Big[ b(x) +Q_{\alpha}(x)\partial^{\alpha}
  + c(x)\partial^2 + 2\big(\Gamma_{\mu}\partial^2
                           +\partial_{\alpha}\Gamma_{\mu}\partial^{\alpha}
                      \big)\partial^{\mu} \Big]
  + O\bigg(\frac{\alpha^2}{\Lambda^4} \bigg),
\end{eqnarray*}
where $\beta = 1+2\alpha\mu^2 / \Lambda^2$ and $\mu$ is a new mass
scale.
It arises as a nonvaninishing
vacuum expectation value of the scalar field $S$, and corresponds to
the constituent quark mass.
The combinations $a(x), b(x), c(x)$ and $Q_{\alpha}(x)$ do not contain
any differential operator acting on the quark fields and they are
defined as
\begin{eqnarray*}
a(x) &=&  i\gamma^{\mu}\big(P_R D_{\mu}\Phi
                            +P_L \overline{D}_{\mu}\Phi^\dagger \big)
        + P_R {\cal M} + P_L {\overline{\cal M}}
        + {1\over 4}[\gamma^\mu ,\gamma^\nu ]\Gamma_{\mu\nu} \,,
\nonumber
\\
b(x) &=& i\gamma^{\mu}\Big[
      P_R \big(A^{(-)}_{\mu} \partial^2\Phi
        + \partial_{\alpha}A^{(-)}_{\mu} \partial^{\alpha}\Phi
        - \Phi \partial^2A^{(+)}_{\mu}
        - \partial_{\alpha}\Phi \partial^{\alpha}A^{(+)}_{\mu} \big)
\nonumber
\\
&&  + P_L \big(A^{(+)}_{\mu} \partial^2\Phi^\dagger
        + \partial_{\alpha}A^{(+)}_{\mu} \partial^{\alpha}\Phi^\dagger
        - \Phi^\dagger \partial^2A^{(-)}_{\mu}
        - \partial_{\alpha}\Phi^\dagger \partial^{\alpha}A^{(-)}_{\mu} \big)
\Big]
\nonumber
\\
&&  + P_R \big(\Phi^\dagger \partial^2\Phi
        + \partial_{\alpha}\Phi^\dagger \partial^{\alpha}\Phi \big)
    + P_L \big(\Phi \partial^2 \Phi^\dagger
        + \partial_{\alpha}\Phi \partial^{\alpha}\Phi^\dagger \big)
    + \Gamma_{\mu} \partial^2\Gamma^{\mu}
\nonumber \\
&&  + \big(\partial_{\alpha}\Gamma_\mu \big)^2
    + {1\over 4}[\gamma^\mu ,\gamma^\nu ] \Big(
      \big[\partial_{\alpha}\Gamma_\mu ,\partial^{\alpha}\Gamma_\nu \big]
      + \Gamma_{\mu} \partial^2\Gamma_{\nu}
      - \Gamma_{\nu} \partial^2\Gamma_{\mu} \Big) \,,
\nonumber
\\
c(x) &=& a(x) + i\gamma^{\mu}\Big[
      P_R \big(A^{(-)}_{\mu}\Phi - \Phi A^{(+)}_{\mu} \big)
    + P_L \big(A^{(+)}_{\mu} \Phi^\dagger - \Phi^\dagger A^{(-)}_{\mu} \big)
\Big]
\nonumber
\\
&&  + P_R {\cal M} + P_L {\overline{\cal M}} + 2\Gamma_{\mu}^2
    + {1\over 4}[\gamma^\mu ,\gamma^\nu ]
                [\Gamma_\mu ,\Gamma_\nu \big] \,,
\nonumber
\\
Q_{\alpha}(x) &=& 3\Gamma_{\mu}\partial_{\alpha}\Gamma^\mu
     + \partial_{\alpha}\Gamma_\mu \,\Gamma^\mu + \partial_{\alpha} a(x)
\nonumber
\\
&&   + 2 i\gamma^{\mu}\Big[
       P_R \big(A^{(-)}_{\mu} \partial_{\alpha}\Phi
               - \Phi \partial_{\alpha}A^{(+)}_{\mu} \big)
     + P_L \big(A^{(+)}_{\mu} \partial_{\alpha}\Phi^\dagger
                - \Phi^\dagger \partial_{\alpha}A^{(-)}_{\mu} \big) \Big]
\nonumber
\\
&&  + 2\big(P_R \Phi^\dagger \partial_{\alpha}\Phi
           + P_L \Phi \partial_{\alpha}\Phi^\dagger \big)
    + {1\over 2}[\gamma^\mu ,\gamma^\nu ] \big(
                 \Gamma_{\mu}\partial_{\alpha}\Gamma_\nu
               - \Gamma_{\nu}\partial_{\alpha}\Gamma_\mu \big) \,.
\end{eqnarray*}
    Here, $\Gamma_{\mu\nu} = \partial_\mu \Gamma_\nu -\partial_\nu \Gamma_\mu
+[\Gamma_\mu ,\Gamma_\nu]\,,$ $\Gamma_\mu=P_L A^{(+)}_\mu + P_R A^{(-)}_\mu\,,$
 ${\cal M} = \Phi^\dagger\Phi - \mu^2\,,$ ${\overline{\cal M}}=\Phi\Phi^\dagger
-
\mu^2\,.$
Furthermore,
$$
D_{\mu}* = \partial_{\mu}* + (A^{(-)}_{\mu}* - *A^{(+)}_{\mu})\,,
\quad
\overline{D}_{\mu}* = \partial_{\mu}* + (A^{(+)}_{\mu}*-*A^{(-)}_{\mu})
$$
are the covariant derivatives; $d_\mu =\partial_\mu +\Gamma_\mu$;
the differential operator $\partial_{\mu}$ acts only on $x$.

%-----------------------------------------------------------------------------
\section{The Effective $p^4$-Lagrangian Including Nonlocal Corrections}
%-----------------------------------------------------------------------------

To obtain the modulus of the quark determinant using the heat-kernel method we
expand
$$
<x\mid \exp (-{\bf \widehat{\widetilde D}}^\dagger{\bf
\widehat{\widetilde D}\tau})\mid y>
$$
around its ``free'' part,
$$
 K_0(x,y;\tau) = <x\mid \exp (-(\beta \hbox{\square }+\mu^{2})\tau)\mid y>
= {1\over (4\pi \beta \tau)^{2}} e^{-\mu^{2}\tau+(x-y)^{2}/(4\beta \tau)}\;,
$$
in powers of the proper-time $\tau$ with the so-called Seeley-deWitt
coefficients $h_{k}(x,y)$,
$$
<x\mid \exp (-{\bf \widehat{\widetilde D}}^\dagger{\bf
\widehat{\widetilde D}\tau})\mid y>
= K_0(x,y;\tau) \sum^{\infty}_{k=0} h_{k}(x,y)\cdot \tau^{k}.
$$
   After integrating over $\tau$ in Eq.\ (\ref{logarithm}) one obtains
$$
{1\over 2}\log (\det {\bf \widehat{D}}^\dagger{\bf \widehat{D}}) =
- {1\over 2} {\mu^{4}\over (4\pi \beta)^{2}} \sum^{\infty}_{k=0}
  {\Gamma (k-2,\mu^{2}/\widetilde{\Lambda}^{2})\over \mu^{2k}}\Trp h_{k}\; ,$$
where
$$
\Gamma (n,x)=\int^{\infty }_{x} d t \, e^{-t}t^{n-1}
$$
is the incomplete gamma function.

    The heat-kernel coefficients $h_k(x) = h_k(x,\;y=x)$ are obtained
from the recursive relation.
\begin{eqnarray*}
&&   2\frac{\alpha}{\Lambda^2}\Gamma_{\mu}t^{\mu}t^2h_{n+3}(x,y)
\nonumber \\ &&
    + \frac{\alpha}{\Lambda^2}\Big[
      t^2 \big(2\Gamma_{\mu}^2+c(x) \big)
    + 2\Gamma_{\mu}\big(3t^{\mu}+2t^{\mu}t_{\alpha}\partial^{\alpha}
                       +t^2\partial^{\mu} \big)
    + 2\partial_{\alpha}\Gamma_{\mu}\,t^{\mu}t^{\alpha} \Big] h_{n+2}(x,y)
\nonumber \\ &&
    + \bigg[ n+1+2t_{\mu}d^{\mu} + \frac{\alpha}{2\Lambda^2} \bigg(
    + 4c(x)\big(1+t_{\alpha}\partial^{\alpha}\big)
    -8\mu^2 t_{\alpha} d^{\alpha} + 2Q_{\alpha}(x)t^{\alpha}
\nonumber \\ &&
    \; + 2\partial_{\alpha}\Gamma_{\mu}
      \big(t^{\mu}\partial^{\alpha} + t^{\alpha}\partial^{\mu} \big)
    + 4\Gamma_{\mu} \big(2\partial^{\mu}
    + 2t_{\alpha}\partial^{\alpha}\partial^{\mu}
    + t^{\mu}\partial^2 \big) \bigg) \bigg] h_{n+1}(x,y)
\nonumber \\ &&
    + \bigg[ a(x) + d_{\mu}d^{\mu} + \frac{\alpha}{\Lambda^2} \bigg(
      b(x) + Q_{\alpha}(x)\partial^{\alpha}
    + 2\big(\Gamma_{\mu}\partial^2
      +\partial_{\alpha}\Gamma_{\mu}\,\partial^{\alpha}\big)\partial^{\mu}
\nonumber \\ &&
    \; + c(x)\partial^2 \bigg)\bigg] h_n(x,y) =0\,,
\end{eqnarray*}
where the differential operator $\partial_\mu$ acts on $x$.
For $V_{\mu}=A_{\mu}=0$, the recursive relation for the heat-kernel
coefficients reduces to
\begin{eqnarray}
&&  \frac{\alpha}{\Lambda^2} t^2 \tilde{c}(x) h_{n+2}(x,y)
\nonumber \\
&&  +\bigg\{ n+1+2t_{\mu}\partial^{\mu} + \frac{\alpha}{\Lambda^2} \bigg[
    2\big(\tilde{a}(x)+\tilde{c}(x)\big) \big(1+t_{\mu}\partial^{\mu}\big)
    -4 \mu^2 t_{\mu} \partial^{\mu}
\nonumber \\
&&  + t_{\mu}\big(\partial^{\mu}\tilde{a}(x)+2\widetilde{Q}^{\mu}(x) \big)
    \bigg] \bigg\} h_{n+1}(x,y)
\nonumber
\\
&&  + \bigg\{ \tilde{a}(x) + \partial^2 + \frac{\alpha}{\Lambda^2} \bigg[
      \tilde{b}(x)
    + \big(\partial_{\mu} \tilde{a}(x)
      +2\widetilde{Q}^{\mu}(x) \big) \partial^{\mu}
\nonumber \\
&&  + \big(\tilde{a}(x)+\tilde{c}(x) \big) \partial^2 \bigg)\bigg] \bigg\}
                                             h_n(x,y) =0\,,
\label{heat-eq}
\end{eqnarray}
   where
$$
 \tilde{a}(x) =  i\gamma^{\mu}\big(P_R \partial_{\mu}\Phi
                                   +P_L \partial_{\mu}\Phi^\dagger \big)
               + P_R {\cal M} + P_L {\overline{\cal M}}\,,
$$
$$
 \tilde{b}(x) = P_R \big(\Phi^\dagger \partial^2\Phi
                         + \partial_{\mu}\Phi^\dagger \partial^{\mu}\Phi \big)
              + P_L \big(\Phi \partial^2 \Phi^\dagger
                         + \partial_{\mu}\Phi \partial^{\mu}\Phi^\dagger
\big)\,,
$$
$$
 \tilde{c}(x) =  P_R {\cal M} + P_L {\overline{\cal M}} \,, \quad
 \widetilde{Q}_{\mu}(x) =   P_R \Phi^\dagger \partial_{\mu}\Phi
                          + P_L \Phi \partial_{\mu}\Phi^\dagger\,.
$$
    The expressions for the heat-coefficients $h_0,...,h_3$ are obtained from
Eq.\ (\ref{heat-eq}) using the computer algebra system REDUCE and the recursive
procedure described in \cite{our},
\begin{eqnarray}
   h_0(x) &=& 1,
\nonumber
\\
\trp[h_1(x)] &=& -\trp \bigg[ \tilde{a} + \frac{\alpha}{\Lambda^2}\bigg(
                \tilde{b} - \tilde{a}\big(\tilde{a}+\tilde{c} \big)
                \bigg) \bigg]+ O\bigg(\frac{\alpha^2}{\Lambda^4} \bigg)\,,
\nonumber
\\
\trp[h_2(x)] &=&  \trp \bigg[ \frac{1}{2}\tilde{a}^2
              + \frac{\alpha}{\Lambda^2} \bigg(\tilde{a}\tilde{b}
              - \frac{2}{3}\tilde{a}^2 \big( \tilde{a}+\tilde{c} \big)
              + \frac{5}{12}\big( \partial_{\mu} \tilde{a} \big)^2
              - \frac{1}{12}\partial_{\mu} \tilde{a}\, \partial^{\mu}\tilde{c}
\nonumber \\ &&
\qquad        - \tilde{a}\, \partial^{\mu}\widetilde{Q}_{\mu}
                \bigg) \bigg] + O\bigg(\frac{\alpha^2}{\Lambda^4} \bigg)\,,
\nonumber
\\
\trp[h_3(x)] &=& -\trp \bigg\{ \frac{1}{6}\tilde{a}^3
              - \frac{1}{12}\big( \partial_{\mu} \tilde{a} \big)^2
              + \frac{\alpha}{\Lambda^2}\bigg[ \frac{1}{2}\tilde{a}^2\tilde{b}
              - \frac{1}{4}\tilde{a}^3 \big( \tilde{a}+\tilde{c} \big)
\nonumber \\ &&
              - \tilde{a}^2
                \bigg( \frac{3}{10} \partial^2 \tilde{a}
                     +\frac{1}{2}\partial^{\mu}\widetilde{Q}_{\mu}
                     +\frac{2}{3}\partial^2 \tilde{c} \bigg)
\nonumber
\\ &&         - \tilde{a} \bigg(
                \frac{5}{6}\big(\partial_{\mu} \tilde{a} \big)^2
              + \frac{1}{15}\partial^2 \tilde{a} \,\tilde{c}
              + \frac{2}{5}\partial_{\mu} \tilde{a}\, \partial^{\mu}\tilde{c}
              + \frac{11}{30}\partial_{\mu} \tilde{c}\, \partial^{\mu}\tilde{a}
\nonumber \\ &&
              + \frac{1}{20} \big( \partial^2 \partial^2 \tilde{c}
                                  +\tilde{c} \partial^2 \tilde{a} \big)
              + \frac{1}{6} \big( \partial^2 \partial^{\mu}\widetilde{Q}_{\mu}
              - \widetilde{Q}_{\mu} \partial^{\mu}\tilde{a}
              + \partial^{\mu}\tilde{a}\,\widetilde{Q}_{\mu}
              - \partial^2 \tilde{b} \big) \bigg)
\nonumber \\ &&
              - \frac{1}{15} \bigg( \partial^2 \tilde{c}\, \partial^2 \tilde{a}
                                  -\tilde{c}\big(\partial_{\mu}\tilde{a}\big)^2
                                  -\big( \partial^2 \tilde{a} \big)^2 \bigg)
              - \frac{1}{18} \mu^2 \big(\partial_{\mu} \tilde{a}\big)^2
                 \bigg] \bigg\}
\nonumber \\ &&
              + O\bigg(\frac{\alpha^2}{\Lambda^4} \bigg)\,.
\label{heat-coeff}
\end{eqnarray}
    First we present the heat-coefficients containing nonlocal
corrections to the minimal (i.e. nonvanishing when $V_{\mu}=A_{\nu}=0)$ part of
the
effective meson lagrangian including $p^2$- and $p^4$-interactions.
    The minimal part has the general form
\begin{eqnarray}
{\cal L}_{eff}^{min}

&=&\frac{F^2_0}{4}\tr\big(\partial_{\mu}U\,\partial^{\mu}U^\dagger\big)
                          +\frac{F_0^2}{4} \tr \big( MU + U^\dagger M \big)
\nonumber \\
&&+ \bigg( L_1-\frac{1}{2} L_2\bigg)\,
    \big( \tr \partial_{\mu}U\,\partial^{\mu}U^\dagger \big)^2
\nonumber \\ &&
  + L_2 \tr \bigg( \frac{1}{2} [\partial_{\mu}U,\partial_{\nu}U^\dagger]^2
  + 3(\partial_{\mu}U\,\partial^{\mu}U^\dagger)^2 \bigg)
\nonumber \\ &&
  + L_3 \tr \big( (\partial_{\mu}U \partial^{\mu}U^\dagger)^2 \big)
  + L_4 \tr \big( \partial_{\mu}U\,\partial^{\mu}U^\dagger \big)\,
        \tr M\big( U + U^\dagger \big)
\nonumber \\ &&
  + L_5 \tr \partial_{\mu}U\,\partial^{\mu}U^\dagger\big( MU +
U^\dagger M \big)\,,
\label{lmin}
\end{eqnarray}
where the  dimensionless structure constants $L_i$ were
introduced by Gasser and Leutwyler in ref.\cite{gasser}.
Since we restrict ourselves to terms of first order in $m_0$, we have
omitted the corresponding higher-order terms from Eq.\ (\ref{lmin}).
Here we have introduced the notations
$$
  U = \exp \left(\frac{i\sqrt{2}}{F_0} \varphi (x) \right)\,,
\quad
  \varphi (x) = \varphi^a(x)\frac{\lambda^a}{2}\,,
$$
where $\varphi (x)$ is the pseudoscalar meson matrix and $F_0$ is the bare
$\pi$ decay constant.

Our calculation predicts the following expression for $F_0$,
\begin{equation}
F_0^2 =   \frac{N_c \mu^2}{4\pi^2} \bigg[ y
        -  \frac{4\pi^2 <\!\!\overline{q}q\!\!>}{\mu^3 N_c}
           \frac{\alpha \mu^2}{\Lambda^2}\bigg]\,,
\label{F0}
\end{equation}
where $y=\Gamma\big(0,\mu^2/\widetilde{\Lambda}^2\big)$.
In  Eq.\ (\ref{F0}) the first term is the standard prediction of the local
limit and
the second corresponds to the nonlocal correction. For the meson mass
matrix $M = diag(\chi^2_u,\chi^2_d,...,\chi^2_n)$ we obtain
\begin{equation}
\chi^2_i = \frac{N_c \mu m^0_i}{2\pi^2 F_0^2} \bigg(
           \widetilde{\Lambda}^2 e^{-\mu^2 / \widetilde{\Lambda}^2} - \mu^2 y
           \bigg)
         =  - \frac{2m^0_i<\!\!\overline{q}q\!\!>}{F^2_0}\,.
\label{Mass}
\end{equation}
    Moreover, the coefficients $L_i$ are given by $L_1-L_2/2=L_4=0$ and
\begin{eqnarray}
  L_2&=& \frac{N_c}{16 \pi^2}\frac{1}{12}\bigg(1
         + 2 \frac{\alpha \mu^2}{\Lambda^2} \bigg)\,,
\nonumber
\\
  L_3&=& -\,\frac{N_c}{16 \pi^2}\frac{1}{6}\bigg(1
          +5(1-y)\frac{\alpha \mu^2}{\Lambda^2} \bigg)\,,
\nonumber
\\
  L_5&=& \frac{N_c}{16 \pi^2} x \bigg[ y-1
         -\frac{28}{3}\frac{\alpha \mu^2}{\Lambda^2} \bigg]\,,
\label{L235}
\end{eqnarray}
where $x = -\mu F_0^2/(2<\!\!\overline{q}q\!\!>)$.
     We made use of the approximation
$\Gamma\big(k,\mu^2/\widetilde{\Lambda}^2\big) \approx \Gamma(k)$
for $k \geq 1$ and $\mu^2 / \widetilde{\Lambda}^2 \ll 1$.

In the local limit, the part of the heat coefficient $h_4$ contributing to the
$p^4$ langrangian is given by
\begin{eqnarray*}
\trp[h_4(x)^{(p^4)}] &=& -\trp \bigg[ \frac{1}{24}\tilde{a}^4
              + \frac{1}{12}\Big( \tilde{a}^2 \partial^2 \tilde{a}
              + \tilde{a} \big(\partial_{\mu} \tilde{a} \big)^2 \Big)
\nonumber \\ &&
              + \frac{1}{720}\Big( 7\big( \partial^2 \tilde{a} \big)^2
              - \big( \partial_{\mu} \partial_{\nu} \tilde{a} \big)^2
                \Big)\bigg].
\end{eqnarray*}
The contributions of $h_4$
linear in $\alpha$ only appear at order $p^6$ which are not
considered here.

     In a completely analogous way we estimate the nonlocal contributions to
the
nonminimal part of the effective meson lagrangian at $O(p^4)$.
     Here we restrict ourselves to the consideration of the structure
coefficients $L_9$ and $L_{10}$ corresponding to the terms
\begin{eqnarray*}
{\cal L}^{(nmin)}_{eff}&=&
    L_9 \tr \Big( F^{(+)}_{\mu \nu}D^{\mu}U\,\overline{D}^{\nu}U^\dagger
                 +F^{(-)}_{\mu \nu}\overline{D}^{\mu}U^\dagger\,D^{\nu}U
            \Big)
\\
&& - L_{10} \tr \Big( U^\dagger F^{(+)}_{\mu \nu}U F^{(-)\mu \nu} \Big)\,,
\end{eqnarray*}
with
\begin{eqnarray*}
  F^{(\pm)}_{\mu \nu} =
       \partial_{\mu} A^{(\pm)}_{\nu}
      -\partial_{\nu} A^{(\pm)}_{\mu}
      +[A^{(\pm)}_{\mu},A^{(\pm)}_{\nu} ]\,.
\end{eqnarray*}
We  obtain
\begin{eqnarray}
  L_9&=& \frac{N_c}{16 \pi^2}\frac{1}{3}\bigg(1
         + \frac{21y-26}{6} \frac{\alpha \mu^2}{\Lambda^2} \bigg)\,,
\nonumber
\\
  L_{10}&=& -\,\frac{N_c}{16 \pi^2}\frac{1}{6}\bigg(1
          +\frac{15y-10}{3} \frac{\alpha \mu^2}{\Lambda^2} \bigg)\,.
\label{L910}
\end{eqnarray}

In Eq.\ (\ref{F0}) we use $<\!\!\overline{q}q\!\!> = - (0.25\,\mbox{GeV})^3$
and $F_0 = 92 \,MeV$ to fix the value of $y$. In comparison with the
local limit it changes from $y_{loc} \sim 1$ to $y_{nonloc}\approx
0.5$. It is worth to be mentioned here that the mass matrix is
not affected by nonlocal corrections.
The nonlocal corrections to the structure constants
$L_2$, $L_3$, $L_9$ and $L_{10}$ (see Eqs.(\ref{L235}) and (\ref{L910})) are
estimated to be not larger than 15-20\% in comparison with their values
in the local limit.
The structure coefficient $L_5$ is most sensitive to nonlocal corrections.
As a result, the splitting of the decay constants $F_\pi$ and $F_K$
\cite{gasser} seems to be strongly influenced by nonlocal effects.

%----------------------------------------------------------------------------
\section{Conclusion}
%----------------------------------------------------------------------------
We have studied the bosonization of an effective
QCD-inspired quark interaction, including nonlocal effects.
First we have considered a "fixed-distance" approximation in order to
obtain a quantitative estimate of the size of the nonlocal
corrections. The result was used as an input into a more general
dynamical separation ansatz which then allowed to predict the
modifications of the structure constants of the chiral meson
lagrangian.
We found that the nonlocal corrections to the $L_i$ coefficients were
typically of the order 15-20\%, except for $L_5$ which is stronly
modified by nonlocal effects.
  Thus, the local NJL model has turned out to be a reasonable approximation for
bosonization of low-energy quark interactions and deriving the effective meson
lagrangian at $O(p^4)$-level. Of course, our analysis could be
extended to any order in the momentum expansion.
Given the fact that nonlocal corrections to the coefficients $L_i$ are of
the order 20 \%, it is to be expected that these
corrections are of the same order of magnitude as local $p^6$
contributions.

The uncertainties arising from nonlocality make it difficult to
distinguish between next to leading order effects of the momentum
expansion and nonlocal contributions to the leading order.
As an example where this is not the case we suggest to investigate the
processes $\eta \to \pi^0 \gamma \gamma$ and $\gamma \gamma \to \pi^0
\pi^0$ where the nonzero {\em Born contributions} to the
amplitudes appear only at $O(p^6)$-level.
\newline
\newline

     The authors are grateful to V.N.Pervushin and M.K.Volkov for useful
discussions and helpful comments.
   One of the authors (A.A.Bel'kov) is grateful for the hospitality
extended to him at TRIUMF, Vancouver. This work was supported in part
by a grant of from the Natural Sciences and Engeneering Research
Council of Canada (S.Scherer) and  by the Russian Grant Center for
Fundamental Researches (A.A.Bel'kov and A.V.Lanyov).

%------------------------------------------------------------------------
%
%                               References
%
%------------------------------------------------------------------------

\end{document}